\documentclass[preprint,showpacs,preprintnumbers,amsmath,amssymb,overcite]{revtex4}
\usepackage{graphicx}
\usepackage{dcolumn}
\usepackage{bm}

\newcommand{\slt}{\!\!\!/}
\newcommand{\sld}{\!\!/}
\newcommand{\beq}{\begin{equation}}
\newcommand{\eeq}{\end{equation}}
\newcommand{\beqa}{\begin{eqnarray}}
\newcommand{\eeqa}{\end{eqnarray}}
\newcommand{\bra}[1]{\mbox{$\langle #1|$}}
\newcommand{\ket}[1]{\mbox{$|#1\rangle$}}

\begin{document}

\title{
{\bf ${\bm K^0}$ photoproduction on the deuteron\\
and the extraction of the elementary amplitude}}

\author{A. Salam\footnote{Permanent address: Departemen Fisika, FMIPA, 
Universitas Indonesia, Depok 16424, Indonesia.} and K. Miyagawa} 
\affiliation{Simulation Science Center, Okayama University of Science, 
1-1 Ridai-cho, Okayama 700-0005, Japan} 

\author{T. Mart}
\affiliation{Departemen Fisika, FMIPA, Universitas Indonesia, 
Depok 16424, Indonesia} 

\author{C. Bennhold}
\affiliation{Center for Nuclear Studies, Department of Physics, 
The George Washington University, Washington, D.C. 20052, USA}

\author{W. Gl\"ockle}
\affiliation{Institut f\"ur Theoretische Physik II, Ruhr-Universit\"at Bochum, 
D-44780 Bochum, Germany}

\date{\today}

\begin{abstract} 
The photoproduction of $K^0$ on the deuteron has been investigated
with the inclusion of $YN$ and $KN$ final-state interaction (FSI), as well as the pion-
mediated process $\gamma d \rightarrow \pi NN \rightarrow KYN$. 
The $YN$ rescattering effects for the inclusive cross section
are found to be large in the threshold regions.
Polarization observables show
sizable FSI effects at larger kaon and hyperon angles.
It is shown that the extraction of the elementary
$ \gamma N \rightarrow KY$  amplitude is
possible in the quasi-free scattering region where FSI effects are
negligible. Furthermore, the cross sections in this region
are  large, indicating that measurements in this kinematical region are favored.

\end{abstract}

\maketitle

\section{Introduction}
\label{introduction}

In order to have a comprehensive understanding of the strong interaction,
it is important to learn more about the hyperon-nucleon ($YN$) interaction. 
It could be studied through hyperon-nucleon scattering in a straightforward manner,
 though the lack of hyperon beams makes it difficult experimentally.
Instead, kaon photoproduction
on the deuteron offers itself as a suitable alternative for studying the hyperon-nucleon
interaction in the final state. Several previous
studies~\cite{ReR67a,ReR67b,AdW89,LiW91} have been done in the inclusive
and exclusive kaon photoproduction on the deuteron
using simple $\Lambda n$ potentials. 
In a recent calculation Yamamura~{\it et al.}~\cite{YaM99} have investigated
$YN$ final state interaction (FSI) effects of the $K^+$ channels using
the more realistic Nijmegen $YN$ potentials~\cite{MaR89,RiS99}. 
Sizeable FSI effects were found in both exclusive and inclusive cross sections,
in particular near the $\Sigma$-threshold. They concluded that precise data
would allow the study of the $YN$ interaction in a great detail. 
This work is extended in Ref.~\cite{Sal03} 
with the inclusion of  kaon-nucleon ($KN$) rescattering in the final state
and the pion-mediated process
$\gamma d \rightarrow \pi NN \rightarrow KYN$ ($\pi N-KY$ process for short)
in the intermediate state. Other recent calculations~\cite{Ker01,Max04} have
also investigated these effects and reached similar conclusions.
        
Regarding  kaon photoproduction on the nucleon, the most
understood channels are the proton
ones~\cite{DaF96,Mar96,BeM96,MaB00,HaC01,JaR02}, i.e., 
$\gamma p\rightarrow K^+\Lambda$ and $\gamma p\rightarrow K^+\Sigma^0$,
since a relatively large number of experimental data are available
for these channels~\cite{Boc94,Tra98,Goers:1999sw}. Meanwhile, the study
of neutron channels is needed in order to complete our understanding
about kaon photoproduction on the nucleon. In these channels the elementary
operator can be quite different. Because of isospin
conservation at the hadronic vertices, there is no $\Lambda$ contribution
in the intermediate states of the $\gamma n \rightarrow K^+\Sigma^-$
process. Furthermore, in the processes
$\gamma n \rightarrow K^0\Lambda$ and $\gamma n \rightarrow K^0\Sigma^0$
there is no $t$-channel contribution in the Born terms since $K^0$ has zero
electrical charge. 
Since free neutron targets are not available to study the neutron channels, 
one needs to use light nuclei like the deuteron or $^3$He as effective
neutron targets. The deuteron is particularly well suited because 
of its small binding energy and its simple structure. Therefore, 
kaon photoproduction on the deuteron is the natural avenue in the
investigation of kaon photoproduction on the neutron. 

With the purpose of extracting the elementary cross section
on the neutron target, Li~{\it et al.}~\cite{LiW92} have calculated
the processes $d(\gamma,K^0p)\Lambda$, $d(\gamma,K^0p)\Sigma^0$, and
$d(\gamma,K^+p)\Sigma^-$, where the kaon is detected in coincidence with
the outgoing nucleon in the impulse approximation (IA). 
They concluded that, within the framework of their model,
 the deuteron can indeed be used to study $K^0$ and $K^+$
 photoproduction from the neutron. 

Very recently an experiment of $K^0$ photoproduction on the deuteron
 has been done at the Laboratory of Nuclear Science (LNS), in 
 Sendai~\cite{Hashimoto}. 
 They measured the cross section of the $d(\gamma,K^0)\Lambda p$ process
 at a photon energy around 1.1 GeV with forward kaon angles. The data are
  now being analyzed. In the near future, they also plan to measure the
  cross section for the exclusive process and some polarization
   observables~\cite{Maeda}. Similarly, for the same reaction
high-precision data from Jefferson Lab are being analyzed that have become available through
the pentaquark searches~\cite{pawel}.
In view of these experiments this paper
  extends the previous work~\cite{YaM99} by calculating  
   $K^0$ photoproduction on the deuteron and taking into account
    the effects of $KN$ rescattering and the $\pi N-KY$ process. 
Other rescattering processes (see, for example, Fig.2
in~\cite{Max04}) will not be included in our study, since
for the kinematics considered here (close to quasi-free scattering 
or to the $YN$ thresholds) they do not contribute significantly.
In extracting the elementary amplitude from the cross section we only
 consider the $\gamma d \rightarrow K^0 \Lambda p$ channel, since the 
 corresponding measurements have been performed. 
 In Sect.~\ref{production-operator}, we briefly review the production
  operator used in this work. The formalism for calculating the transition
   matrix and observables are shown in
    Sect.~\ref{reaction-on-deuteron}. 
        The results are presented in Sect.~\ref{results} and we close the
         paper with conclusions in Sect.~\ref{conclusions}.

\section{The kaon photoproduction operator}
\label{production-operator}

Most analyses of kaon photoproduction on the nucleon have been performed 
at tree level in an effective Lagrangian approach. While this leads to
violation of
unitarity, this kind of isobar model provides a simple tool to parameterize
the elementary process because it is relatively easy to calculate
and to use for production on nuclei. In this approach, the photoproduction
amplitude $ \gamma N \rightarrow KY $ ( in the following denoted as
elementary amplitude) can be written as
\begin{eqnarray}
\langle\vec p_{Y}\mu_{Y}
\vert t^{\gamma K}_{\lambda}\vert
\vec p_{N}\mu_{N}\rangle
&=& 
\bar{u}_{\mu_{Y}}
\Big(\sum_{i=1}^{4} A_{i} \Gamma^{i}_{\lambda}\Big)
u_{\mu_{N}}\,,~~
\label{eq-gnky-amplitude}
\end{eqnarray}
where $A_{i}$'s are invariant amplitudes as functions of the Mandelstam
variables only. The hyperon and nucleon Dirac spinors are denoted
by $u_{\mu_{Y}}$ and $u_{\mu_{N}}$, respectively. The invariant Dirac
operators $\Gamma^{i}_{\lambda}$, which are given by
\begin{eqnarray}
\Gamma^{1}_{\lambda} 
& = & {\textstyle \frac{1}{2}} \gamma_{5} 
\left({\epsilon\sld}\!_{\lambda} k \slt 
- k \slt {\epsilon\sld}\!_{\lambda} \right)\,, \\
\Gamma^{2}_{\lambda} 
& = & \gamma_{5} \left[ (2q-k) \cdot \epsilon_{\lambda} P \cdot k
- (2q-k) \cdot k P \cdot \epsilon_{\lambda} \right]\,, \\
\Gamma^{3}_{\lambda} 
& = & \gamma_{5} \left( q\cdot k {\epsilon\sld}\!_{\lambda}
- q \cdot \epsilon_{\lambda} k \slt \right)\,, \\
\Gamma^{4}_{\lambda} 
& = & i \epsilon_{\mu \nu \rho \sigma} \gamma^{\mu} q^{\nu}
\epsilon^{\rho}_{\lambda} k^{\sigma}\,,
\label{eq-gnky-Gamma-matrix}
\end{eqnarray}
are gauge invariant Lorentz pseudoscalars and given in terms of the usual
$\gamma$-matrices, the photon momentum $k$, and its polarization vector
$\epsilon_{\lambda}$. Here $\lambda$ labels the polarization states, 
$q$ the meson momentum, and $P=(p'+p)/2$, where $p$ and $p'$ denote
initial and final baryon momenta, respectively~\cite{Don72}.
By expressing the Dirac operators and spinors in term of Pauli matrices
$\vec\sigma$ and spinors $\chi$, we can write the kaon photoproduction
operator as
\begin{eqnarray}
t^{\gamma K}_{\lambda} 
&=& i\Big(L_{\lambda}+i\vec\sigma\cdot\vec K_{\lambda}\Big)\,,
\label{eq-gnky-LK-form}
\end{eqnarray}
where $L_{\lambda}$ and $\vec K_{\lambda}$ are functions of $A_{i}$.
The rather lengthy expression of $A_{i}$, $L_{\lambda}$, and
$\vec K_{\lambda}$ can be found in Ref.~\cite{Mar96}.
\begin{figure}[htbp]
\begin{center}
\includegraphics[width=.64\textwidth]{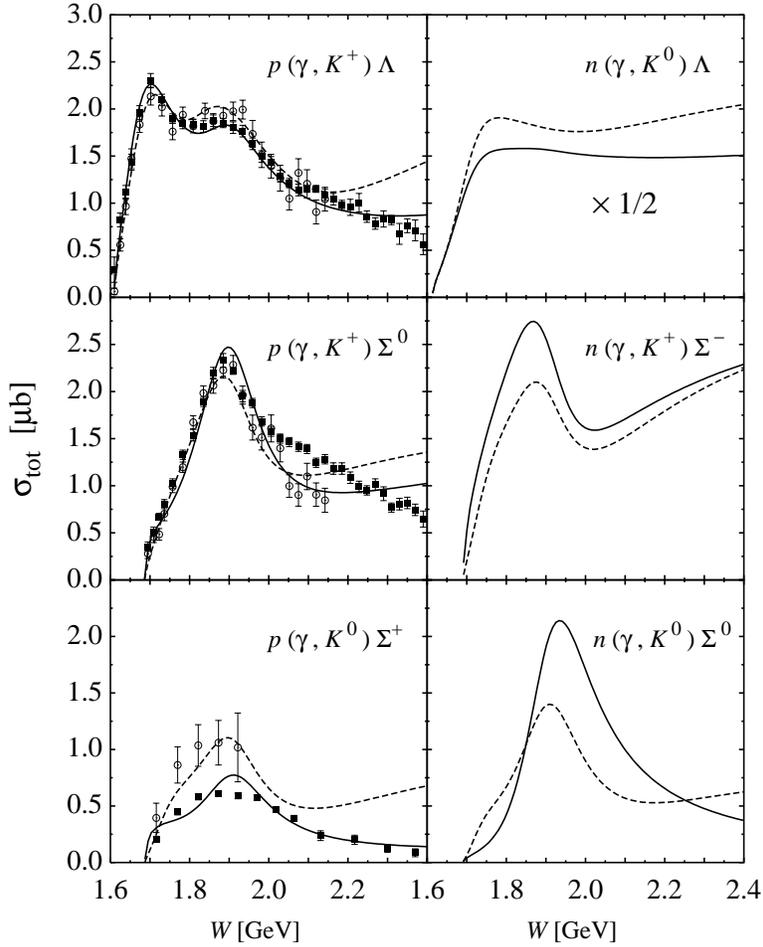}
\end{center}
\caption{Total cross section of kaon photoproduction on the nucleon obtained 
by using the original KAON-MAID model (dashed lines) and the same model but
refitted to the new SAPHIR data (solid lines). 
Older experimental data are taken from Refs.~\cite{Tra98,Goers:1999sw}, while
new SAPHIR data are taken from Refs. \cite{Lawall:2005np,Glan03b}.} 
\label{fig-gnky-total-cross-section}
\end{figure}

In the present work we use the KAON-MAID model~\cite{MaB00}, which includes 
the $D_{13}$(1895) resonance beside the Born terms and other resonances. 
Separate hadronic form factors for each vertex were used. In order to restore gauge 
invariance, the recipe from Haberzettl~\cite{Hab97} was utilized. 
The coupling constants and cut-off parameters were determined by fitting to 
the experimental data. The behavior of this model in all six isospin channels 
is exhibited by the dashed lines in Fig.~\ref{fig-gnky-total-cross-section}. 

Recently, the SAPHIR collaboration at ELSA has published new data on all
proton channels, i.e. the $\gamma p\rightarrow K^+\Lambda$,
$\gamma p\rightarrow K^+\Sigma^0$ and $\gamma p\rightarrow K^0\Sigma^+$ ones.
The new data are more precise and cover all angular distributions
in the energy range from threshold up to $W\simeq 2.4$ GeV. Furthermore, as
shown by the middle- and lower-left panels of 
Fig.~\ref{fig-gnky-total-cross-section}, 
these new data show a significant discrepancy with the previous experiment
in the 
$\gamma p\rightarrow K^+\Sigma^0$ and $\gamma p\rightarrow K^0\Sigma^+$
channels.
Since KAON-MAID was fitted to previous data, it obviously faces a problem
to reproduce the new ones. Figure \ref{fig-gnky-total-cross-section}
reveals this problem explicitly. 

To investigate the effects of new data on KAON-MAID we refit the coupling
constants
in this isobar model by only using the new SAPHIR data in our database.
The results
are shown by the solid lines in Fig.~\ref{fig-gnky-total-cross-section}. 
While 
in the $\gamma p\rightarrow K^+\Lambda$ and $\gamma p\rightarrow K^0\Sigma^+$
channels KAON-MAID can nicely reproduce the new measurements, it is 
obviously unable to explain the $\gamma p\rightarrow K^+\Sigma^0$ cross
section
at $W\gtrsim 2.0$ GeV. This problem originates from the fact that KAON-MAID
does not have certain resonances at this energy region. By floating the
resonance
masses during the fit process, a recent study has shown that the new data on
this 
channel demand a nucleon resonance $S_{11}$ with $M\simeq 2100$
MeV \cite{Lawall:2005np}.

The predicted total cross sections for neutron channels are given in the
three right panels of Fig.~\ref{fig-gnky-total-cross-section}. 
There are sizable differences between the original prediction of KAON-MAID
and the refitted version. Nevertheless, all predicted cross sections are of the same
order.

\begin{figure}[t]
\begin{center} 
\includegraphics[width=0.65\textwidth]{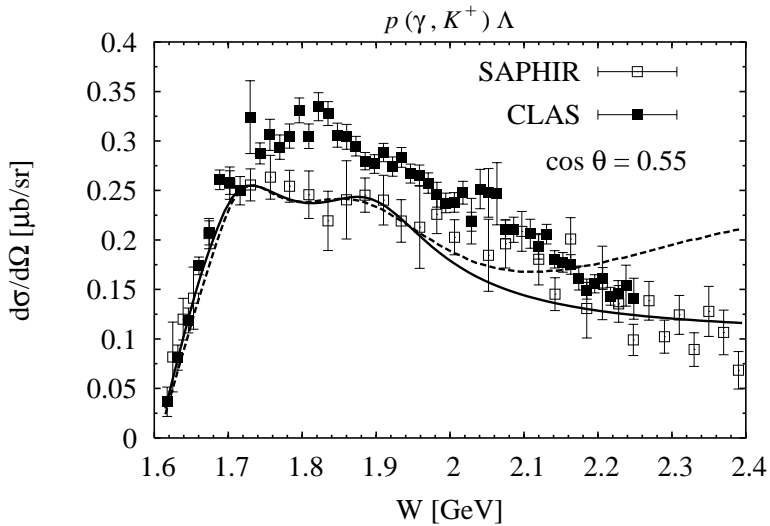} 
\end{center}
\caption{Example of the systematic discrepancy between SAPHIR (open squares) 
  \cite{Lawall:2005np}
  and CLAS (solid squares) \cite{McNabb:2003nf} 
 data in the differential cross section of the $\gamma p\rightarrow K^+\Lambda$ 
  channel as a function of the total c.m. energy $W$. 
  Predictions from the two elementary operator models given in 
  Fig.~\ref{fig-gnky-total-cross-section} are shown for comparison.} 
\label{fig-elem-dif1} 
\end{figure}

Very recently, the CLAS collaboration has also published their data for 
both $\gamma p\rightarrow K^+\Lambda$ and $\gamma p\rightarrow K^+\Sigma^0$
channels \cite{McNabb:2003nf,Bradford:2005pt}. The CLAS data show, 
unfortunately, substantial and
systematic discrepancies with the SAPHIR ones. This problem is clearly
illustrated
in Fig.~\ref{fig-elem-dif1}. As a consequence, an effort
to simultaneously fit the model to both data versions would be meaningless.
In view of this, we
decide to use the original KAON-MAID model in the subsequent calculations.
This choice is also supported by the fact that in this paper most calculations on
the deuteron have been performed at low photon energy, a region
where the original and the refitted KAON-MAID models are still in good
agreement and
the discrepancy in the new experimental data is not too significant. 
Furthermore,
our main motivation in this paper is to show the possibility to extract the
elementary
cross section for the $ \gamma N \rightarrow KY $ process from the
deuteron cross section. Therefore, we can expect that the main 
conclusion will be independent from the choice of elementary model.

\section{The reaction on the deuteron}
\label{reaction-on-deuteron}
\subsection{Cross sections and amplitudes}
\label{cross-section-amplitude}

The general expression for the cross section
using the convention of Bjorken and Drell~\cite{BjDr} 
is given by
\begin{eqnarray}
d\sigma &=&
\frac{1}{|\vec v_{\gamma}-\vec v_{d}|}
\frac{m_Y m_N}{8E_{\gamma}E_{K} E_{d} E_Y E_N} 
\frac{d\vec p_{K}}{(2\pi)^3} 
\frac{d\vec p_{Y}}{(2\pi)^3} 
\frac{d\vec p_{N}}{(2\pi)^3}\,
(2\pi)^{4}\delta^{4}(P_{d}+Q-p_{Y}-p_{N})\,
\nonumber\\ &&\times\,
\frac{1}{6} \sum_{\mu_{Y}\mu_{N}\mu_{d}\lambda} 
\big\vert\sqrt{2}
%
%
\langle\vec p_{Y}\vec p_{N}\mu_{Y}\mu_{N}
\vert T^{\gamma K}_\lambda\vert
\Psi_{\mu_{d}}\rangle\,
\big\vert^2\,,
\label{eq-gdkyn-general-cross-section}
\end{eqnarray}
where $\mu_{Y}$, $\mu_{N}$, $\mu_{d}$, and $\lambda$ denote 
the spin projections of hyperon, nucleon, deuteron and the photon 
polarization, respectively.
For the deuteron state $\Psi_{\mu_{d}}$, however, 
we use a noncovariant notation,
which removes the standard additional factor
$
1/2E_{d}(2\pi)^3
$.
 Here $Q=p_{\gamma}-p_{K}$ is the momentum transfer
to the two-baryon system and the factor $\sqrt{2}$ comes from the proper
antisymmetrization. 
In this expression the dependencies on the kinematical variables have been
suppressed and it should be understood that
the production operator acts on one of
the 
two initial baryons. 
By integrating the right hand side of Eq.~(\ref{eq-gdkyn-general-cross-section})
over the three-momentum of nucleon and momentum of hyperon,
followed by rewriting the flux factor,
we arrive at the cross section for the exclusive process $d(\gamma,KY)N$ 
in the deuteron rest frame, i.e.,
\begin{eqnarray}
\frac{d\sigma}{dp_{K}d\Omega_{K}d\Omega_{Y}} &=&
\frac{m_{Y}m_{N}\vert\vec p_{K}\vert^2\vert\vec p_{Y}\vert^2}
{4(2\pi)^2E_{\gamma}E_{K}}\,
\big\vert (E_Y+E_N)\vert\vec p_{Y}\vert 
-E_{Y}\vec Q\cdot\hat p_{Y}\big\vert^{-1}\,
\nonumber\\ &&\times\,
\frac{1}{6} \sum_{\mu_{Y}\mu_{N}\mu_{d}\lambda} 
\big\vert\sqrt{2}
\langle\vec p_{Y}\vec p_{N}\mu_{Y}\mu_{N}
\vert T^{\gamma K}_\lambda\vert
\Psi_{\mu_{d}}\rangle\big\vert^2\, .
\label{eq-gdkyn-exclusive-cross-section}
\end{eqnarray}
Through out the paper we work in the deuteron rest frame. 
For the inclusive process $d(\gamma,K)YN$ the cross section 
is given by
\begin{eqnarray}
\frac{d\sigma}{dp_{K}d\Omega_{K}} &=&  
\int d\Omega^{\,\rm cm}_{Y}\, 
\frac{m_{Y}m_{N}\vert\vec p_{K}\vert^2\vert\vec p^{\,\rm cm}_{Y}\vert}
{4(2\pi)^2E_{\gamma}E_{K}W}\,
\nonumber\\ &&\times\,
\frac{1}{6} \sum_{\mu_{Y}\mu_{N}\mu_{d}\lambda} 
\big\vert\sqrt{2}
\langle\vec p_{Y}\vec p_{N}\mu_{Y}\mu_{N}
\vert T^{\gamma K}_\lambda\vert
\Psi_{\mu_{d}}\rangle\,
\big\vert^2\,,
\label{eq-gdkyn-inclusive-cross-section}
\end{eqnarray}
where $W^2=(P_{d}+Q)^2$ and $\vert\vec p^{\,\rm cm}_{Y}\vert$ is the hyperon
momentum 
calculated in the center of mass frame of the two final baryons.

\begin{figure}[t]
\begin{center}
\includegraphics[width=.5\textwidth]{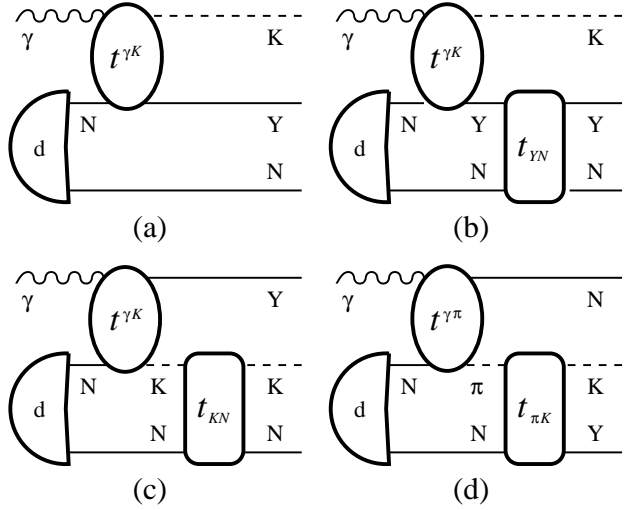}
\end{center}
\caption{\small Kaon photoproduction on the deuteron. 
The diagrams indicate (a) the impulse approximation (IA), 
(b) the $YN$ rescattering, 
(c)  the $KN$ rescattering, 
and (d) the $\pi N-KY$ process.}
\label{fig-gdkyn-diagram}
\end{figure}

The amplitude is approximated by the diagram shown in 
Fig.~\ref{fig-gdkyn-diagram}, 
which is written for convenience as
\begin{eqnarray}
T^{\gamma K}_\lambda &=& t^{\gamma K}
+ t^{\gamma K}_{YN}
+ t^{\gamma K}_{KN}
+ t^{\gamma K}_{K\pi}\,, 
\label{eq-gdkyn-operator}
\end{eqnarray}
where $t^{\gamma K}$, $t^{\gamma K}_{YN}$, $t^{\gamma K}_{KN}$, 
and $t^{\gamma K}_{K\pi}$ denote the operators for the impulse approximation, 
hyperon-nucleon rescattering, kaon-nucleon rescattering, 
and the pion mediated process, respectively.

The IA and $YN$ rescattering terms are calculated 
precisely. Here we describe the calculation briefly; 
the reader should consult  Ref.~\cite{YaM99} for details. 
The operator of the sum of diagrams (a) and (b) can be written as 
\begin{eqnarray}
T^{\gamma K}_{YN} &=& t^{\gamma K} + t^{\gamma K}_{YN} \\
&=& t^{\gamma K} + t_{YN}\, G_{YN} t^{\gamma K}\,
\label{eq-gdkyn-yn-operator}
\end{eqnarray}
with the $YN$ scattering operator $t_{YN}$ which obeys
the Lippmann-Schwinger equation 
\begin{eqnarray}
t_{YN} &=& V_{YN} + V_{YN}\, G_{YN}\, t_{YN}\,, 
\label{eq-gdkyn-yn-lippmann-schwinger}
\end{eqnarray}
where $V_{YN}$ denotes the $YN$ potential operator 
and $G_{YN}$ the free $YN$ propagator. 
Inserting Eq.~(\ref{eq-gdkyn-yn-lippmann-schwinger}) 
into Eq.~(\ref{eq-gdkyn-yn-operator}), we obtain 
\begin{eqnarray}
T^{\gamma K}_{YN} &=& t^{\gamma K} + V_{YN}\, 
G_{YN}\, T^{\gamma K}_{YN}\,, 
\label{eq-gdkyn-yn-operator-V}
\end{eqnarray}
which can be solved by inversion, i.e.,
\begin{eqnarray}
T^{\gamma K}_{YN} &=& 
\left(1 - V_{YN} \, G_{YN} \right)^{-1}
t^{\gamma K}\,. 
\label{eq-gdkyn-yn-inversion}
\end{eqnarray}
After solving the last equation in a partial wave decomposition with
respect to the $YN$ subsystem, one
obtains the $YN$ rescattering amplitude by subtraction of the IA term.


The $KN$ rescattering (diagram (c) in Fig.~\ref{fig-gdkyn-diagram})
is evaluated 
directly in contrast to $YN$ rescattering. The corresponding operator
is given by 
\begin{eqnarray}
t^{\gamma K}_{KN} &=& t_{KN}\, G_{KN}\, t^{\gamma K}\,,
\label{eq-gdkyn-kn-operator}
\end{eqnarray}
where $t_{KN}$ is the $KN$ scattering operator,
which also obeys the Lippmann-Schwinger equation 
of the form~(\ref{eq-gdkyn-yn-lippmann-schwinger}), 
and $G_{KN}$ is the free $KN$ propagator. 
For $V_{KN}$ we take a rank-1 separable potential which, 
in the partial wave representation, is given by
\begin{eqnarray}
V_{\ell J}(p^{\prime},p) &=& \lambda_{\ell J}\, 
g(p^{\prime})_{\ell J}\,g(p)_{\ell J}
\label{eq-knkn-separable-potential}
\end{eqnarray}
with the form factor 
\begin{eqnarray}
g(p)_{\ell J} &=& \frac{B_{\ell J}\, p^{\ell}}
{\left[p^{2}+A_{\ell J}^{2}\right]^{\frac{\ell+2}{2}}}\,,
\label{eq-knkn-form-factor}
\end{eqnarray}
where $\lambda_{\ell J}$, $B_{\ell J}$, and $A_{\ell J}$ are parameters
which are 
determined by fitting the phase shift to the experimental
data~\cite{HyA92,Sax80}. 
The $\pi N-KY$ process (diagram (d) in Fig.~\ref{fig-gdkyn-diagram}) is 
calculated in the same fashion; 
details of these calculations can be found in Ref.~\cite{Sal03}.

\subsection{Polarization Observables}
\label{polarization-observable}

With respect to polarization observables, we consider the tensor
target asymmetries $T_{2M}$ which are given by~\cite{Are88}
\begin{eqnarray}
T_{2M} \frac{d\sigma}{d\Omega_{K}} &=&
(2-\delta_{M0})\, {\mathcal Re}\, V_{2M}\,,\quad M=0, 1, 2\,,
\label{eq-gdkyn-T2m}
\end{eqnarray}
where
\begin{eqnarray}
V_{2M} &=& 
\sqrt{15} 
\sum_{\mu_{Y}\mu_{N}\lambda} 
\sum_{\mu^{\prime}_{d}\mu_{d}}
(-1)^{1-\mu^{\prime}_{d}}
\left(\begin{array}{ccc}
    1       &      1                &  2 \\
\mu_{d} & -\mu^{\prime}_{d} & -M 
\end{array}\right) 
\nonumber\\ &&\times\,
\int_{p_K^{\rm min}}^{p_K^{\rm max}} dp_{K} \int d\Omega^{\,\rm cm}_{Y}\,\kappa \,
{\cal M}_{\mu_{Y}\mu_{N}\mu_{d}\lambda}^{*}
{\cal M}_{\mu_{Y}\mu_{N}\mu^{\prime}_{d}\lambda}
\label{eq-gdkyn-V2m}
\end{eqnarray}
with a kinematic factor
\begin{eqnarray}
\kappa &=& 
\frac{m_{Y}m_{N}\vert\vec p_{K}\vert^{2}\vert\vec p^{\,\rm cm}_{Y}\vert}
{24(2\pi)^{2} E_{\gamma}E_{K} W}\,.
\label{eq-gdkyn-kinematic-factor}
\end{eqnarray}
We also calculate the hyperon recoil polarization $P_{y}$, 
the beam asymmetry $\Sigma$, and the double polarization $C_{x}$ and $C_{z}$, 
which are given by
\begin{eqnarray}
P_{y} &=&
\frac{{\rm Tr}\,{\cal M}{\cal M}^{+}\sigma_{y}}
{{\rm Tr}\,{\cal M}{\cal M}^{+}}\,,
\\
\Sigma &=& 
\frac{{\rm Tr}\,{\cal M}_{\epsilon_{y}}{\cal M}^{+}_{\epsilon_{y}}
-{\rm Tr}\,{\cal M}_{\epsilon_{x}}{\cal M}^{+}_{\epsilon_{x}}}
{{\rm Tr}\,{\cal M}_{\epsilon_{y}}{\cal M}^{+}_{\epsilon_{y}}
+{\rm Tr}\,{\cal M}_{\epsilon_{x}}{\cal M}^{+}_{\epsilon_{x}}}\,,
\\
C_{x} &=&
\frac{{\rm Tr}\,{\cal M}_{\epsilon_{1}}{\cal M}^{+}_{\epsilon_{1}}\sigma_{x}}
{{\rm Tr}\,{\cal M}_{\epsilon_{1}}{\cal M}_{\epsilon_{1}}}\,,
\\
C_{z} &=&
\frac{{\rm Tr}\,{\cal M}_{\epsilon_{1}}{\cal M}^{+}_{\epsilon_{1}}\sigma_{z}}
{{\rm Tr}\,{\cal M}_{\epsilon_{1}}{\cal M}_{\epsilon_{1}}}\,.
\label{eq-gdkyn-polarization-observable}
\end{eqnarray}
In these equations the amplitude ${\cal M}$ reads 
\begin{eqnarray}
\cal M &=&
\sqrt{2}\,
\langle\vec p_{Y}\vec p_{N}\mu_{Y}\mu_{N}
\vert T^{\gamma K}_\lambda\vert
\Psi_{\mu_{d}}\rangle\,  ,
\label{eq-gdkyn-amplitude-M}
\end{eqnarray}
where, for example, ${\cal M}_{\epsilon_{y}}$ indicates the amplitude 
with photon polarization pointing into the y-axis, i.e.
$\vec\epsilon_{\lambda}$ = $\vec\epsilon_{y}$. 
The photon polarization
$\vec\epsilon_{1}$ = $-\frac{1}{\sqrt{2}}(\vec\epsilon_{x}
+i\vec\epsilon_{y})$ describes the helicity state +1. 
The beam asymmetry $\Sigma$ is obtained with linearly polarized photon, 
while the double polarization observables $C_{x}$ and $C_{z}$ 
are the observables for both polarized hyperon and circularly polarized photon.

\subsection{Extraction of the elementary amplitude}
\label{extraction-elementary-amplitude}

In the following, we describe briefly the extraction of the invariant 
elementary amplitude from the cross section on the deuteron 
in view of the study of kaon photoproduction on the neutron. 
The condition for the extraction must be that FSI effects
be negligible. 
Thus, the amplitude can be approximated only by the impulse term.  
In this case the invariant elementary amplitude can be separated  
from the deuteron one. To show this, first 
we write the amplitude of the impulse 
term explicitly as
\begin{eqnarray}
\bra{\vec p_{Y}\vec p_{N}\mu_{Y}\mu_{N}}
t^{\gamma K}_{\lambda}
\ket{\Psi_{\mu_{d}}} &=& 
\sum_{\mu_{N'}}\bra{\vec p_{Y}\mu_{Y}} 
t^{\gamma K}_{\lambda}
\ket{-\vec p_{N}\mu_{N'}}\,
C^{\frac{1}{2}\,\frac{1}{2}\,1}_{\mu_{N}\mu_{N'}\mu_{S}} 
\Psi_{\mu_{S}\mu_{d}}(-\vec p_{N})\,,
\label{eq-gdkyn-explicit-impulse-amplitude}
\end{eqnarray}  
where $C$ denotes Clebsch-Gordan coefficient
and $\vec p_{N}$ is the spectator nucleon momentum. 
The deuteron wave function $\Psi$ is given by
\begin{eqnarray}
\Psi_{\mu_{S}\mu_{d}}(-\vec p_{N}) &=& \sum_{\ell}  
C^{\ell\,1\,1}_{\mu_{\ell}\mu_{S}\mu_{d}}\, u_{\ell}(\vert\vec p_{N}\vert)\,
Y_{\ell\mu_{\ell}}(-\hat p_{N})\,,
\label{eq-gdkyn-deuteron-wave-function}
\end{eqnarray}
where $Y$ indicates spherical harmonics and $u$ is its radial part, 
which is generated by the Nijmegen93 NN potential~\cite{nij93} in this work. 
Summing over all spin states of the amplitude 
in Eq.~(\ref{eq-gdkyn-explicit-impulse-amplitude}), after some algebra 
we find 
\begin{eqnarray}
\sum_{\mu_{Y}\mu_{N}\mu_{d}\lambda} 
\big\vert
\bra{\vec p_{Y}\vec p_{N}\mu_{Y}\mu_{N}}
t^{\gamma K}_{\lambda} 
\ket{\Psi_{\mu_{d}}}\big\vert^2 &=&
D\,\sum_{\mu_{Y}\mu_{N'}\lambda}
\big\vert\bra{\vec p_{Y}\mu_{Y}} 
t^{\gamma K}_{\lambda} 
\ket{-\vec p_{N}\mu_{N'}}\big\vert^2\,,
\label{eq-gdkyn-separation-elementary-amplitude}
\end{eqnarray}
where 
\begin{eqnarray}
D &=& 
{\textstyle\frac{3}{2}}\vert Y_{00}\vert^2 u_{0}^2 
+\left(
{\textstyle\frac{3}{10}}\vert Y_{20}\vert^2 +
{\textstyle\frac{3}{5}}\vert Y_{21}\vert^2 +
{\textstyle\frac{3}{5}}\vert Y_{22}\vert^2
\right) u_{2}^2\,.
\label{eq-gdkyn-deuteron-factor}
\end{eqnarray}
With this expression we can write
Eq.~(\ref{eq-gdkyn-exclusive-cross-section}) as
\begin{eqnarray}
\frac{d\sigma}{dp_{K}d\Omega_{K}d\Omega_{Y}} &=&
\frac{m_{Y}m_{N}\vert\vec p_{K}\vert^2\vert\vec p_{Y}\vert^2}
{4(2\pi)^2E_{\gamma}E_{K}}\,
\big\vert (E_Y+E_N)\vert\vec p_{Y}\vert 
-E_{Y}\vec Q\cdot\hat p_{Y}\big\vert^{-1}\,
\nonumber\\ &&\times\,
\frac{1}{6}\, D\sum_{\mu_{Y}\mu_{N'}\lambda}
\big\vert\bra{\vec p_{Y}\mu_{Y}} 
t^{\gamma K}_{\lambda} 
\ket{-\vec p_{N}\mu_{N'}}\big\vert^2\,.
\label{eq-gdkyn-exclusive-cross-section-extract}
\end{eqnarray}
On the r.h.s. of this equation the sum of the squared amplitudes 
of the elementary process $\gamma N \rightarrow KY$ has been completely 
separated from the deuteron wave function.

\section{Results}
\label{results}

\begin{center}
\begin{figure}[htbp]
\hfill
\begin{minipage}[t]{.46\textwidth} 
\begin{center} 
\includegraphics[width=1.\textwidth]{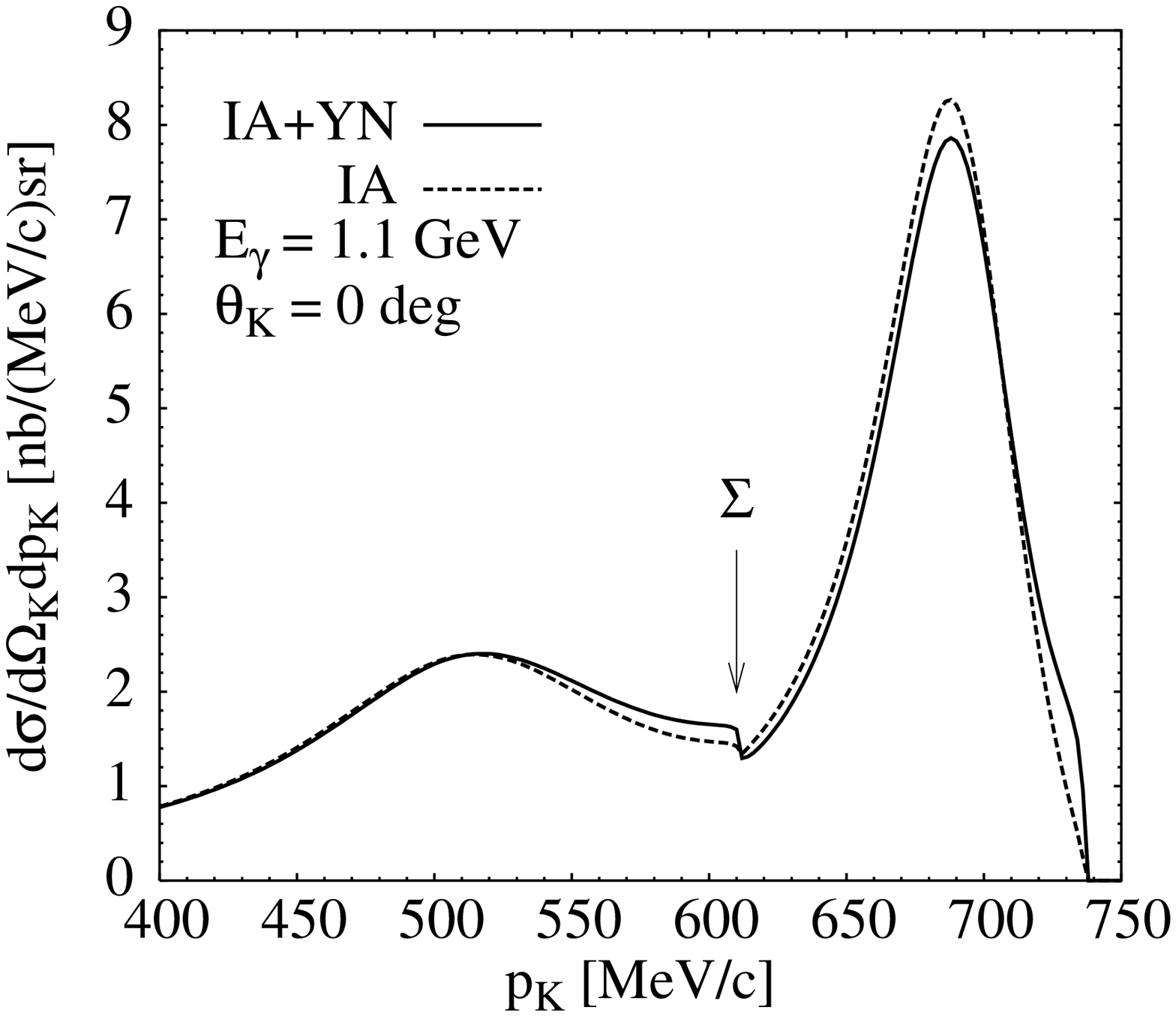} 
\end{center}
\caption{
Inclusive cross sections of $d(\gamma,K^0)YN$ as functions
of $p_{K}$ at $E_{\gamma}=1.1$ GeV and $\theta_{K}= 0^\circ$. 
The dashed line denotes the IA results and the  solid line
  the IA+$YN$ results. 
The arrow indicates the $\Sigma$ threshold.} 
\label{fig-gdkyn-inclusive-fsi} 
\end{minipage} 
\hfill
\begin{minipage}[t]{.46\textwidth} 
\begin{center}
\includegraphics[width=1.\textwidth]{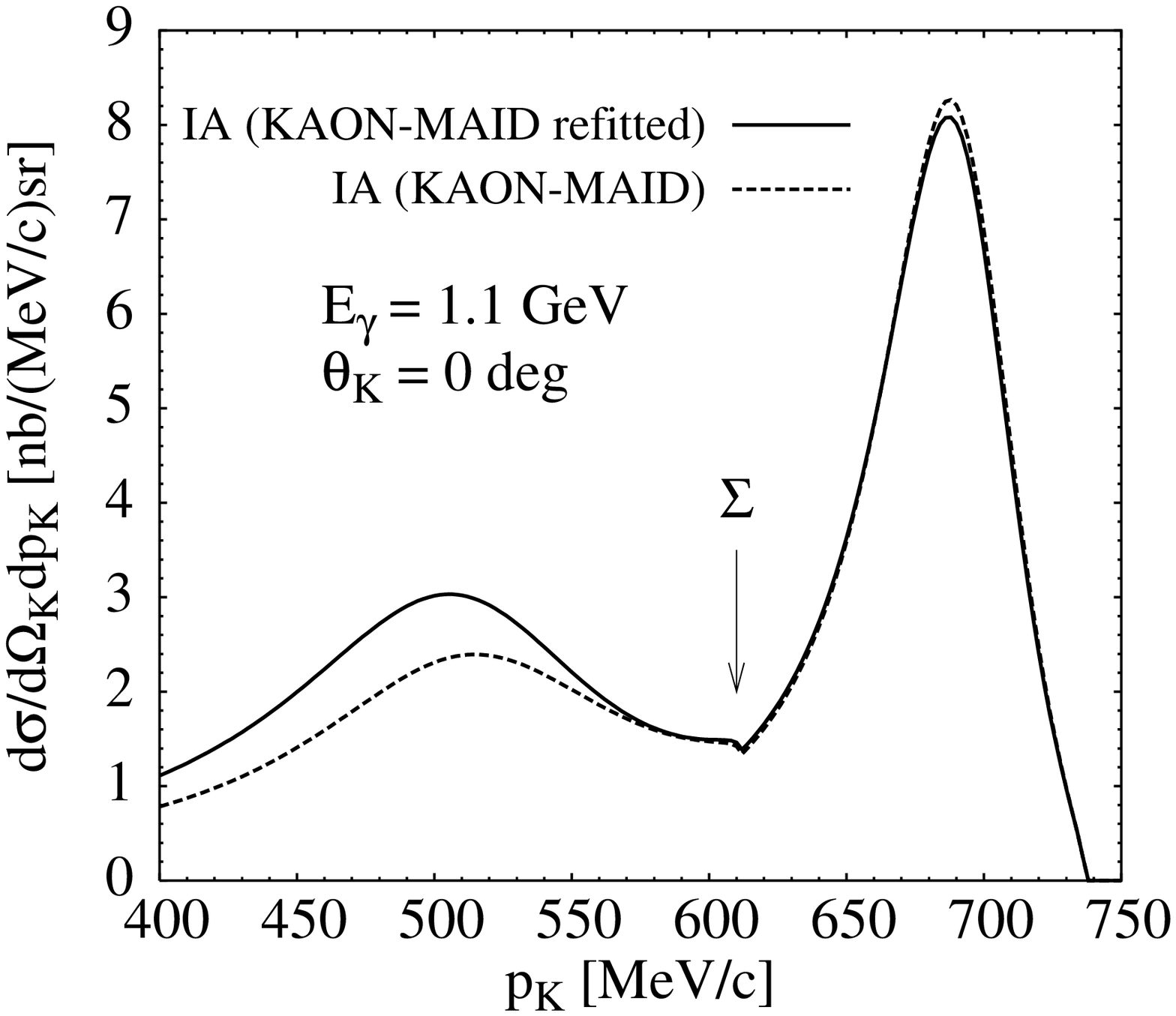} 
\end{center}
\caption{Comparison of the inclusive cross sections
in IA
using the two different elementary models
given in Fig~\ref{fig-gnky-total-cross-section}. 
}
\label{fig-gdkyn-inclusive-ia} 
\end{minipage}
\end{figure} 
\end{center}

The inclusive cross section, calculated 
using Eq.~(\ref{eq-gdkyn-inclusive-cross-section}) 
at a photon energy of $E_{\gamma}$ = 1.1 GeV and at forward kaon angle
$\theta_{K} = 0^\circ$, 
is shown in Fig.~\ref{fig-gdkyn-inclusive-fsi}. 
In this figure we add up the cross section for
all outgoing channels, 
$K^0\Lambda p$, $K^0\Sigma^0 p$ and $K^0\Sigma^+ n$.  
The positions of the two peaks are found to be consistent with the
$\Lambda$ and $\Sigma$ quasi-free scattering (QFS) conditions. 
We note sizable effects of $YN$-rescattering in the 
$\Lambda$ and $\Sigma$ threshold regions, and on the top of the
$ \Lambda$-peak around $p_K=690$ MeV/$c$. 
Figure~\ref{fig-gdkyn-inclusive-ia} shows 
the inclusive cross sections in IA, where the results for
the two different elementary operators discussed 
in Sec.~\ref{production-operator} are compared. This figure 
shows a relatively large variation of the cross sections
around the $\Sigma$-QFS region. This result originates from
the different models for the elementary process
(see Fig.~\ref{fig-gnky-total-cross-section})
 of the $K^+\Sigma^-$ channel at this energy of $W\simeq 1.7$ GeV.
 The other $K\Sigma$ channels also exhibit significant differences,
but since their overall cross sections are  smaller compared to $K^+\Sigma^-$,
their effects become less important. In contrast to the $ K^+ \Sigma^- $
channel,
for both the $K^+\Lambda$ and $K^0\Lambda$ channels
 the corresponding cross sections do not differ too much around 
$W=1.7$ GeV. This also explains
why the difference between the two elementary models does not show
up strongly around the $\Lambda$-QFS region.

By comparing Figs. ~\ref{fig-gdkyn-inclusive-fsi} 
and \ref{fig-gdkyn-inclusive-ia}, we can see that the variation 
originating from the different elementary operators is larger than 
that originating from the FSI effects around the $\Sigma$-QFS region. 
In other words, we can say that the cross section of the process 
$d(\gamma,K^0)\Sigma N$ is more sensitive to the choice of the 
elementary operators than to the FSI effects. This feature shows 
that the process $d(\gamma,K^0)\Sigma N$ can serve as a way to access 
the elementary operators. 
Experimental data with error bars smaller than these variations are needed now
to apply our procedure.


\begin{figure}[t]
\hfill
\begin{minipage}[b]{.44\textwidth} 
\begin{center}
\includegraphics[width=1.\textwidth]{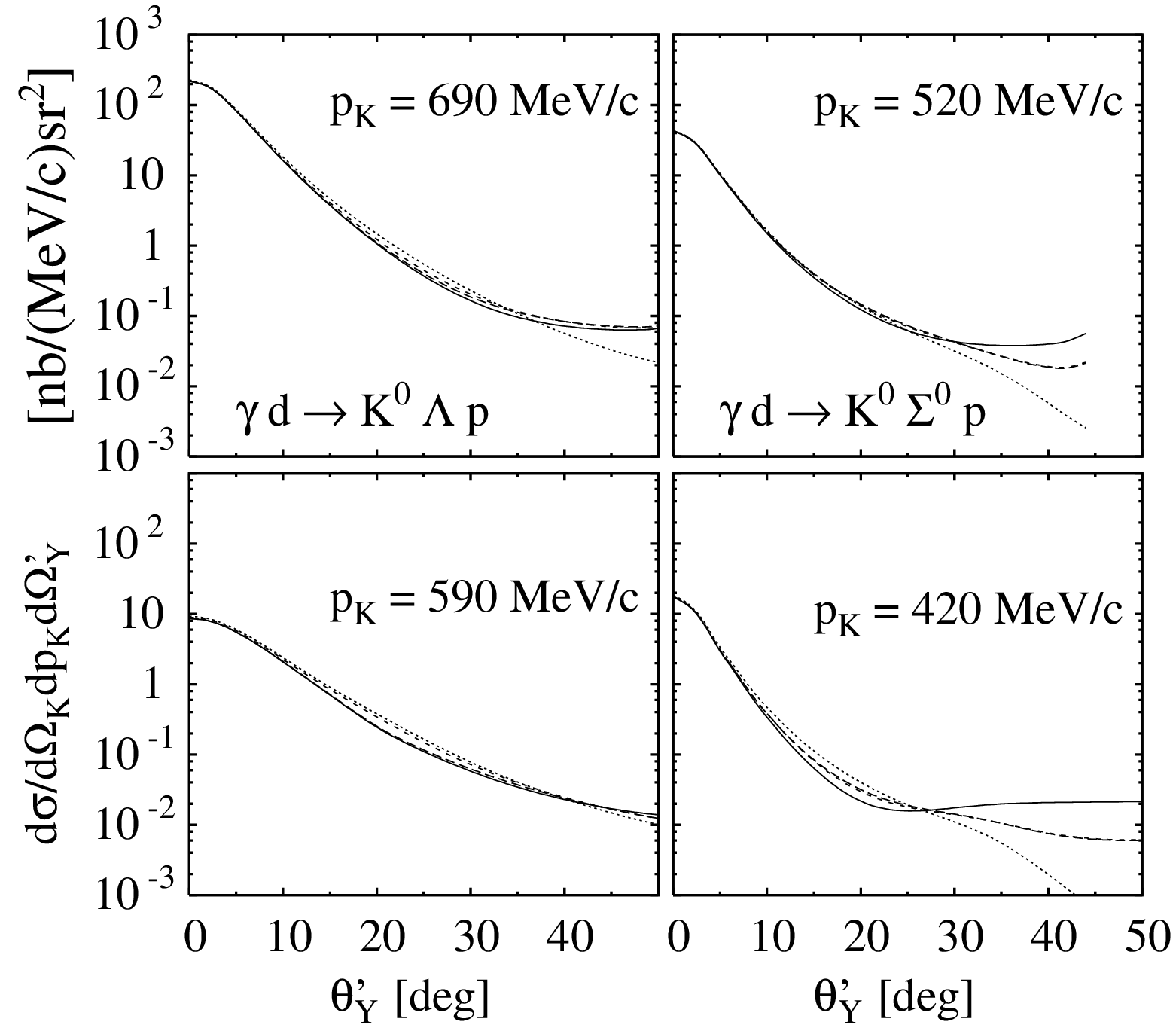}
\end{center}
\caption{Exclusive cross sections 
as a function of hyperon angle $\theta_{Y}$ 
for $\gamma d \rightarrow K^0\Lambda p$ (left panels), 
and $\gamma d \rightarrow K^0\Sigma^0 p$ (right panels) 
at the photon energy $E_{\gamma}$ = 1.1 GeV, 
kaon angle $\theta_{K}=0^\circ$, 
hyperon angle $\phi_{Y} = 0^\circ$ 
for two values of the kaon momentum $\vec p_{K}$. 
The dotted line corresponds to IA, the short-dashed line to IA+$YN$, 
the dashed line to  IA+$YN$+$KN$, and the solid line
to  IA+$YN$+$KN$+($\pi N-KY$).}
\label{fig-exclusive-cross-section}
\end{minipage} 
\hfill
\begin{minipage}[b]{.44\textwidth} 
\begin{center}
\includegraphics[width=1.\textwidth]{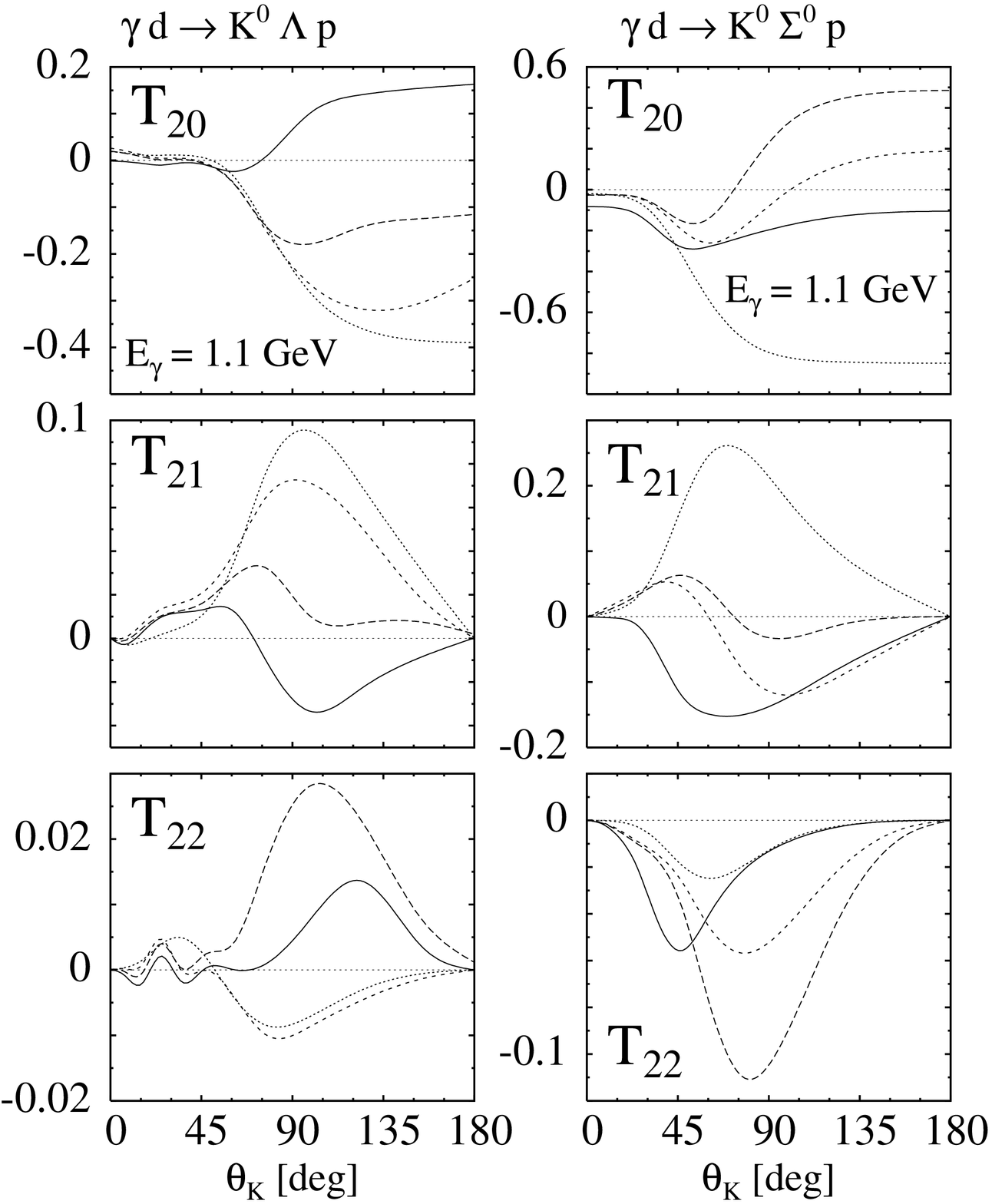}
\end{center}
\caption{Tensor asymmetries $T_{20}$, $T_{21}$, and $T_{22}$, 
as a function of kaon angle $\theta_{K}$ 
for $\gamma d \rightarrow K^{0}\Lambda p$ (left panels), 
and $\gamma d \rightarrow K^{0}\Sigma^{0} p$ (right panels) 
for the photon energy $E_{\gamma}$ = 1.1 GeV. 
The lines are as in 
Fig.~\ref{fig-exclusive-cross-section}. 
}
\label{fig-tensor-asymmetry-1100}
\end{minipage}
\end{figure} 
Figure~\ref{fig-exclusive-cross-section} presents the exclusive cross
sections 
calculated for the photon energy 1.1 GeV at forward kaon angle
$\theta_{K} = 0^\circ$. 
The figure shows that the effect of the $YN$ rescattering is stronger 
at larger hyperon angles, 
while the effects of $KN$ rescattering and $\pi N-KY$ process
are negligible or small.

The tensor asymmetries for the photon energy
of $E_{\gamma}$ = 1.1 GeV 
are shown in Fig.~\ref{fig-tensor-asymmetry-1100}. 
This figure shows that 
the $YN$, $KN$ rescatterings, and $\pi N-KY$ process have different
effects on $T_{20}$, $T_{21}$, and $T_{22}$ at larger kaon angles. 
\begin{figure}[h]
\begin{center}
\includegraphics[width=.75\textwidth]{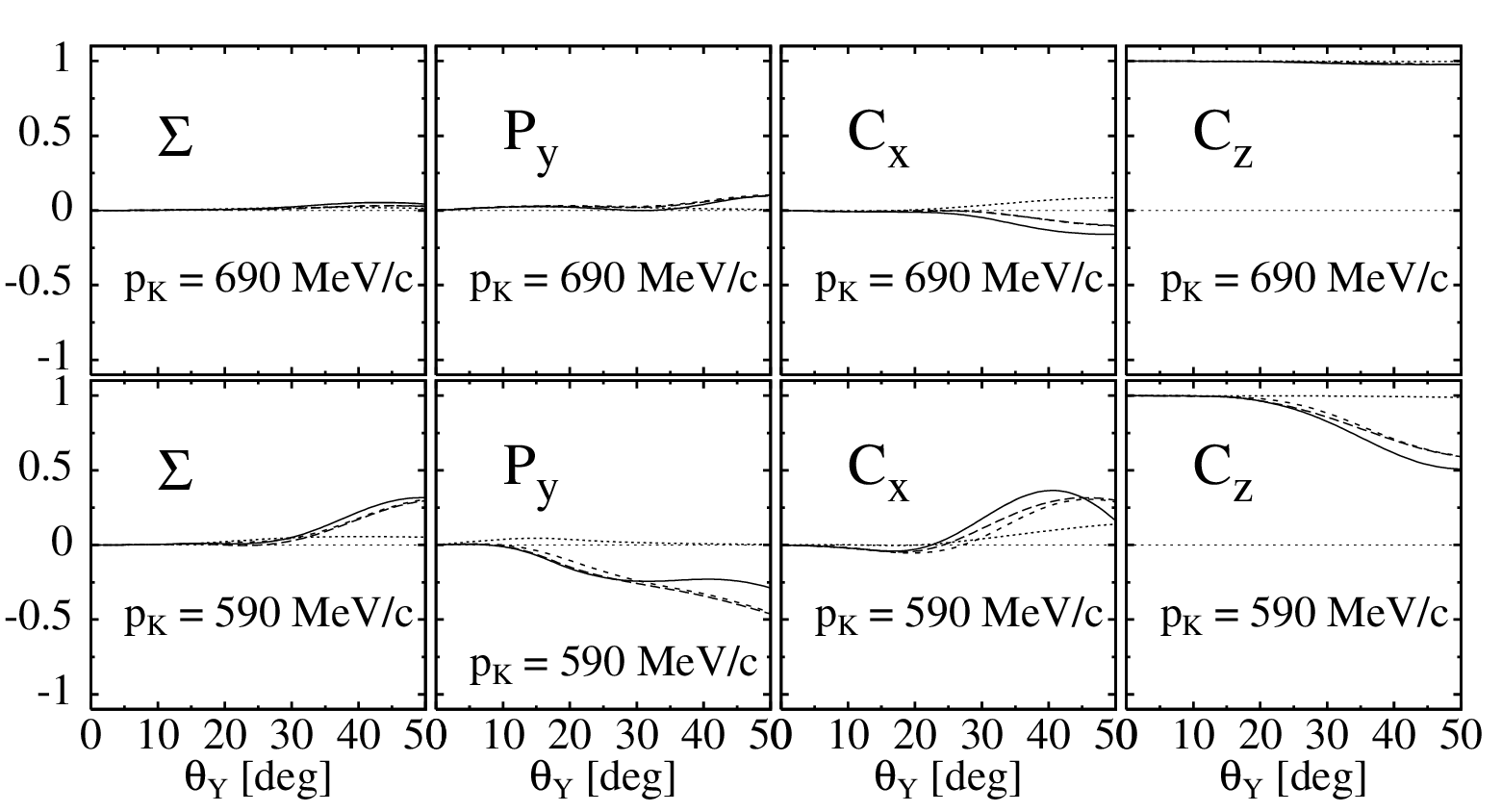}
\end{center}
\caption{Beam asymmetry $\Sigma$, recoil polarization $P_y$, 
and double polarizations $C_{x}$ and $C_{z}$, 
as functions of the hyperon angle $\theta_{Y}$  
for the $\gamma d \rightarrow K^0\Lambda p$
process. 
Top and bottom panels are for the two kaon momenta
$p_K=690$ and 590 MeV/$c$, respectively.
The photon energy is $E_{\gamma}$ = 1.1 GeV, 
the kaon angle is fixed
at  $\theta_{K} = 0^\circ$
and the hyperon azimuthal angle is $\phi_{Y} = 0^\circ$.
The lines are as in 
Fig.~\ref{fig-exclusive-cross-section}.
}
\label{fig-polarization-observable-24}
\end{figure} 
\begin{figure}[!h]
\begin{center}
\includegraphics[width=.75\textwidth]{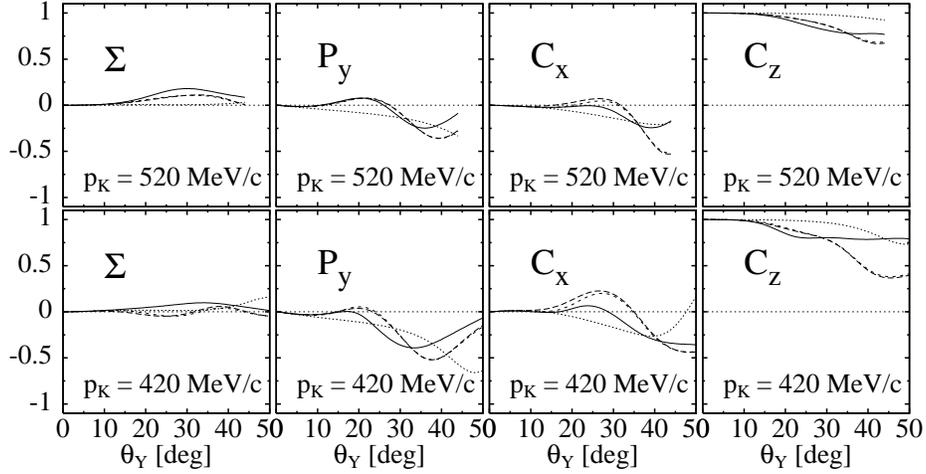}
\end{center}
\caption{Same as in Fig.~\ref{fig-polarization-observable-24} 
but for $\gamma d \rightarrow K^0\Sigma^0 p$ with
$p_K=520$ and 420 MeV/$c$.
}
\label{fig-polarization-observable-64}
\end{figure} 
Figures~\ref{fig-polarization-observable-24} and
\ref{fig-polarization-observable-64} 
show the polarization observables
$\Sigma$, $P_{y}$, $C_{x}$, and $C_{z}$ in the 
$\gamma d \rightarrow K^0\Lambda p$ and 
$\gamma d \rightarrow K^0\Sigma^{0} p$ channels, respectively. 
These polarization observables are calculated for the photon energy
$E_{\gamma}$ = 1.1 GeV, 
the kaon angle $\theta_{K} = 0^\circ$, 
and the hyperon angle $\phi_{Y} = 0^\circ$. 
The top panels correspond to the results at the peak positions 
of the inclusive cross section, 
while the bottom panels refer to  lower kaon momenta. For these observables
the $YN$
rescattering effects 
dominate at larger hyperon angles $\theta_{Y}$. The effects are 
more remarkable at lower kaon momenta
than at the peak positions.

\begin{figure}[t]
\hfill
\begin{minipage}[t]{.44\textwidth} 
\begin{center}
\includegraphics[width=1.\textwidth]{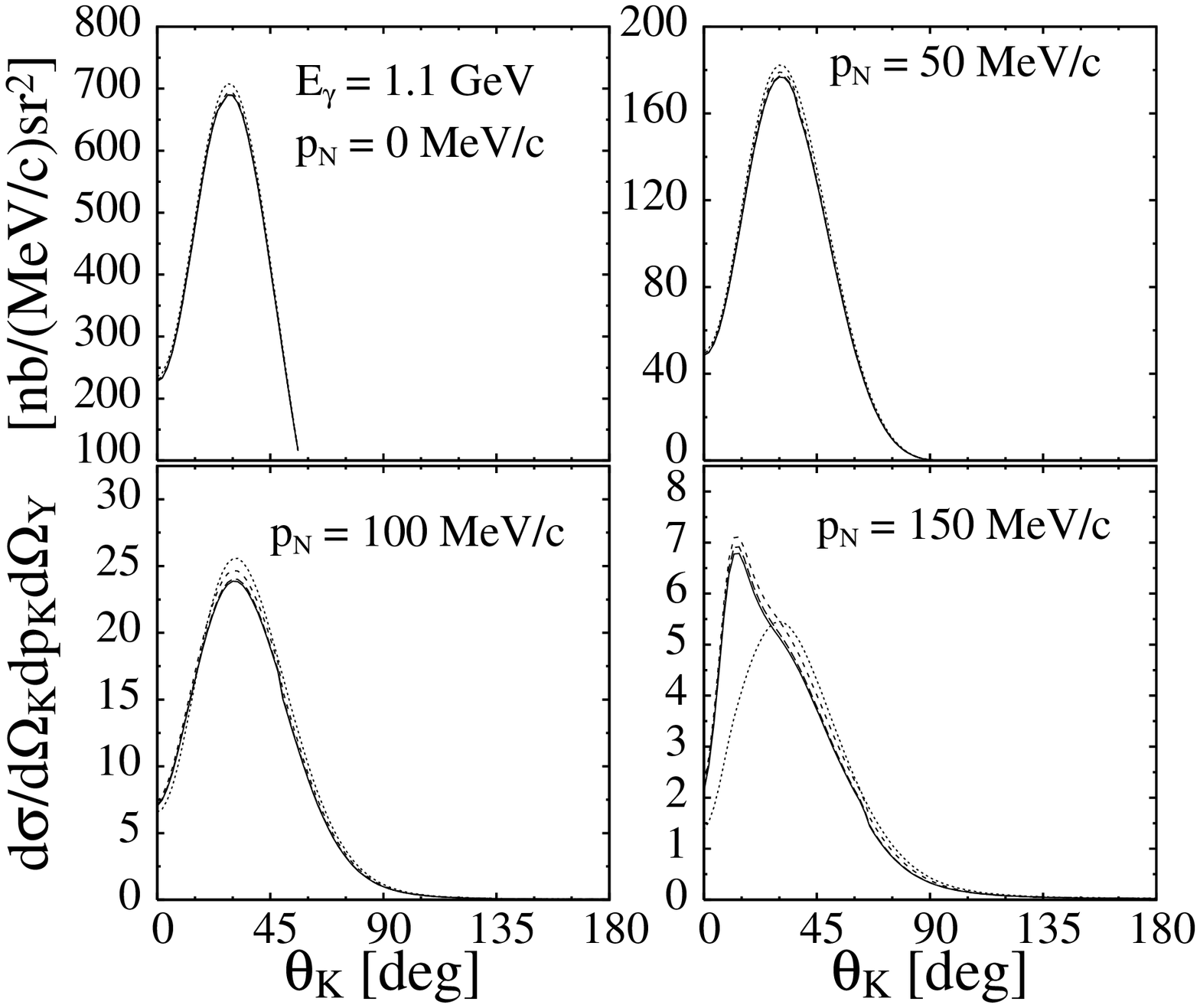}
\end{center}
\caption{Exclusive cross sections of $d(\gamma,K^0\Lambda)p$ 
as functions of kaon angle $\theta_{K}$ 
for the photon energy $E_{\gamma}$ = 1.1 GeV. 
The direction of the outgoing nucleon momentum $\vec p_{N}$ is fixed
at $\theta_{N} = 30^\circ$ and $\phi_{N} = 180^\circ$, but 
the magnitude of $\vec p_{N}$ is varied. The results
for $p_{N}=0$, 50, 100, and 150 MeV/$c$ are shown.
The lines are as in
Fig.~\ref{fig-exclusive-cross-section}.
}
\label{fig-qf-exclusive-cross-section-1100}\end{minipage} 
\hfill
\begin{minipage}[t]{.44\textwidth} 
\begin{center}
\includegraphics[width=1.\textwidth]{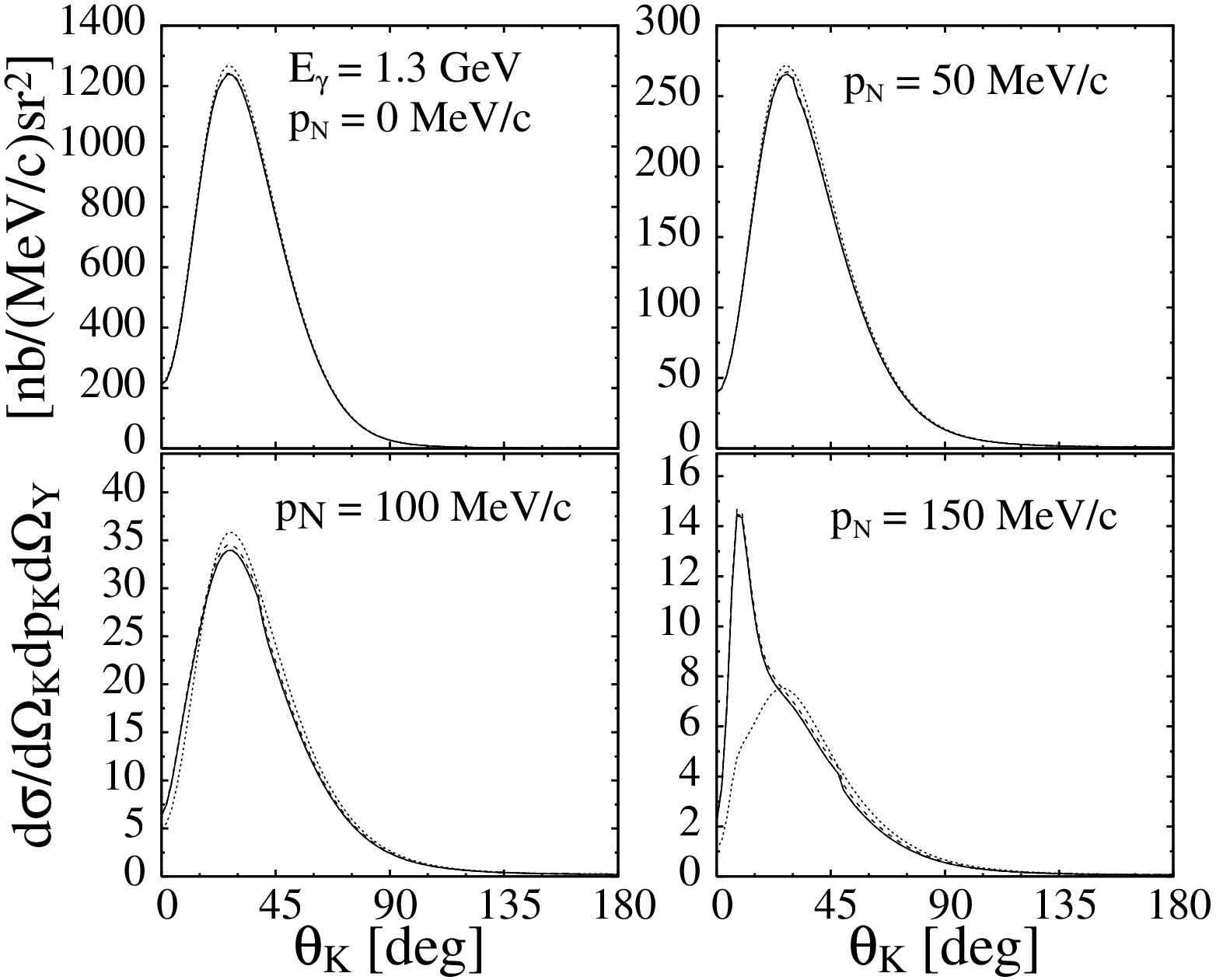}
\end{center}
\caption{Same as Fig.~\ref{fig-qf-exclusive-cross-section-1100} 
but for the photon energy $E_{\gamma}$ = 1.3 GeV.}
\label{fig-qf-exclusive-cross-section-1300}\end{minipage}
\end{figure} 

\begin{figure}[t]
\begin{center}
\includegraphics[width=.5\textwidth]{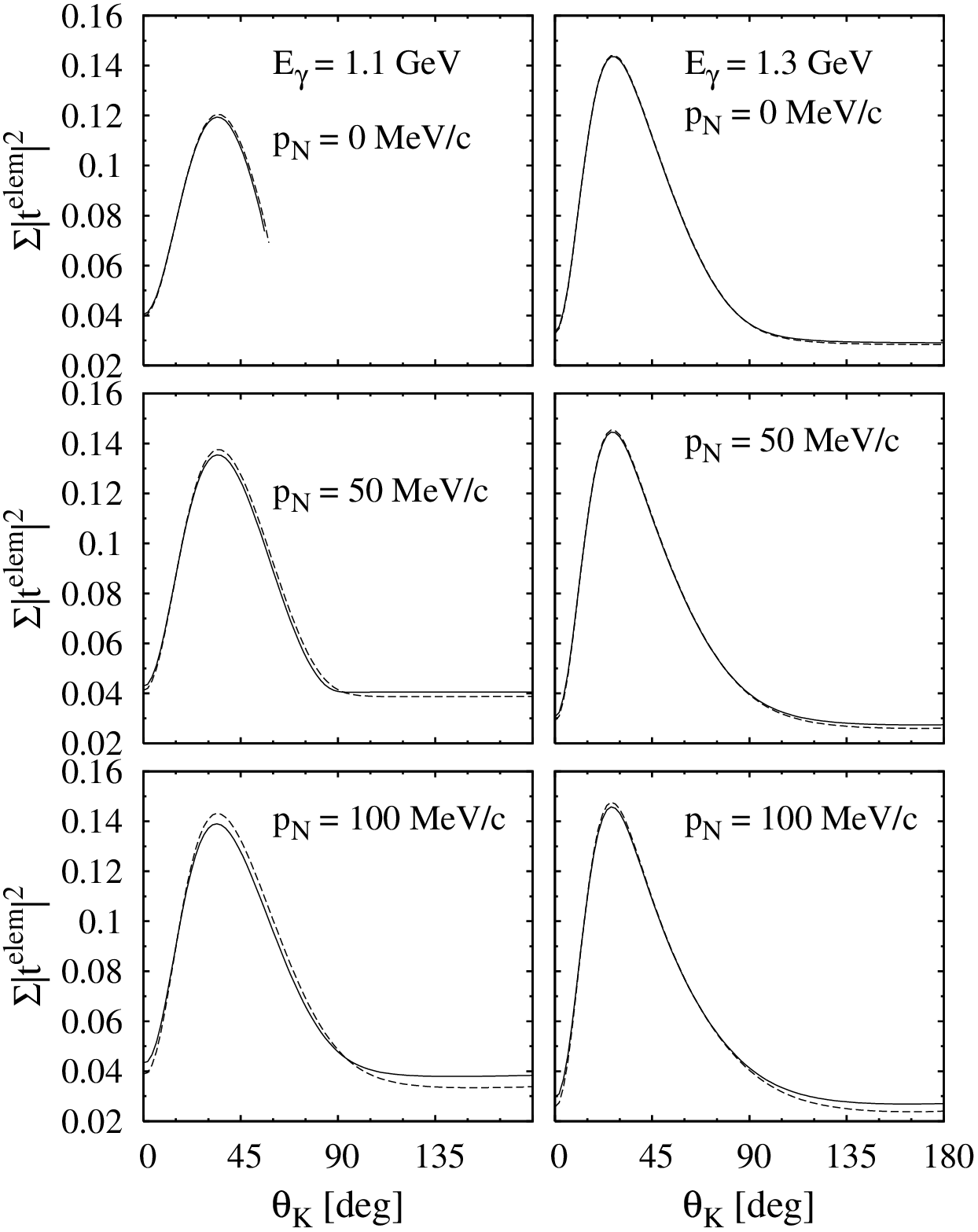}
\end{center}
\caption{Comparison between the extracted and the free-process 
amplitudes squared 
for the process $\gamma n \rightarrow K^{0}\Lambda$
at the photon energy $E_{\gamma}$ = 1.1 GeV (left panels)
and $E_{\gamma}$ = 1.3 GeV (right panels).
The solid line refers to  the extracted amplitudes
from $\gamma d \rightarrow K^{0}\Lambda p$, 
while the dashed line is obtained from the free-process amplitudes.
The direction of $\vec p_{N}$ is fixed at $\theta_{N} = 30^\circ$
and $\phi_{N} = 180^\circ$, but 
the magnitude of $\vec p_{N}$ is varied with 0, 50, 100
150 MeV/$c$.
}
\label{fig-extraction-elementary-amplitude}
\end{figure}

Figures \ref{fig-qf-exclusive-cross-section-1100} and 
\ref{fig-qf-exclusive-cross-section-1300} 
show the exclusive cross sections of $d(\gamma,K^0\Lambda)p$ 
as a function of kaon angle $\theta_{K}$ 
for the photon energies $E_{\gamma}$ = 1.1 and 1.3 GeV, respectively. 
In the figures we fix
the direction of the outgoing nucleon momentum $\vec p_{N}$ to 
$\theta_{N} = 30^\circ$ and $\phi_{N} = 180^\circ$, 
while  several magnitudes of $\vec p_{N}$, namely 0, 50, 100, and
150 MeV/$c$ are studied. 
At nucleon momentum $p_{N}=150$ MeV/$c$,  
the effects of $YN$ rescattering are most prominent
at forward kaon angles,
but almost negligible for zero nucleon momentum, i.e. 
under QFS kinematics.
From these features we conclude that the extraction 
of the elementary amplitude is favored in QFS kinematics.

Finally, in Fig.~\ref{fig-extraction-elementary-amplitude}
we compare the extracted elementary amplitudes for
$p_{N}=0$, 50, 100 MeV/$c$ with the corresponding 
free-process amplitudes.
We see that the extracted amplitudes are in good agreement with
the free-process amplitudes at QFS kinematics. 
At $p_{N}=50$ MeV/$c$ there is a small discrepancy between 
the extracted and the free-process amplitudes, 
especially at larger kaon angles. 
The discrepancy grows at $p_{N}=100$ MeV/$c$. 
One can also note from the figures that
the extraction works better done at higher 
photon energies $E_{\gamma}$
where the discrepancy is smaller for the same nucleon momentum
$p_{N}$.  
\begin{figure}[htbp]
\begin{center}
\includegraphics[width=.7\textwidth]{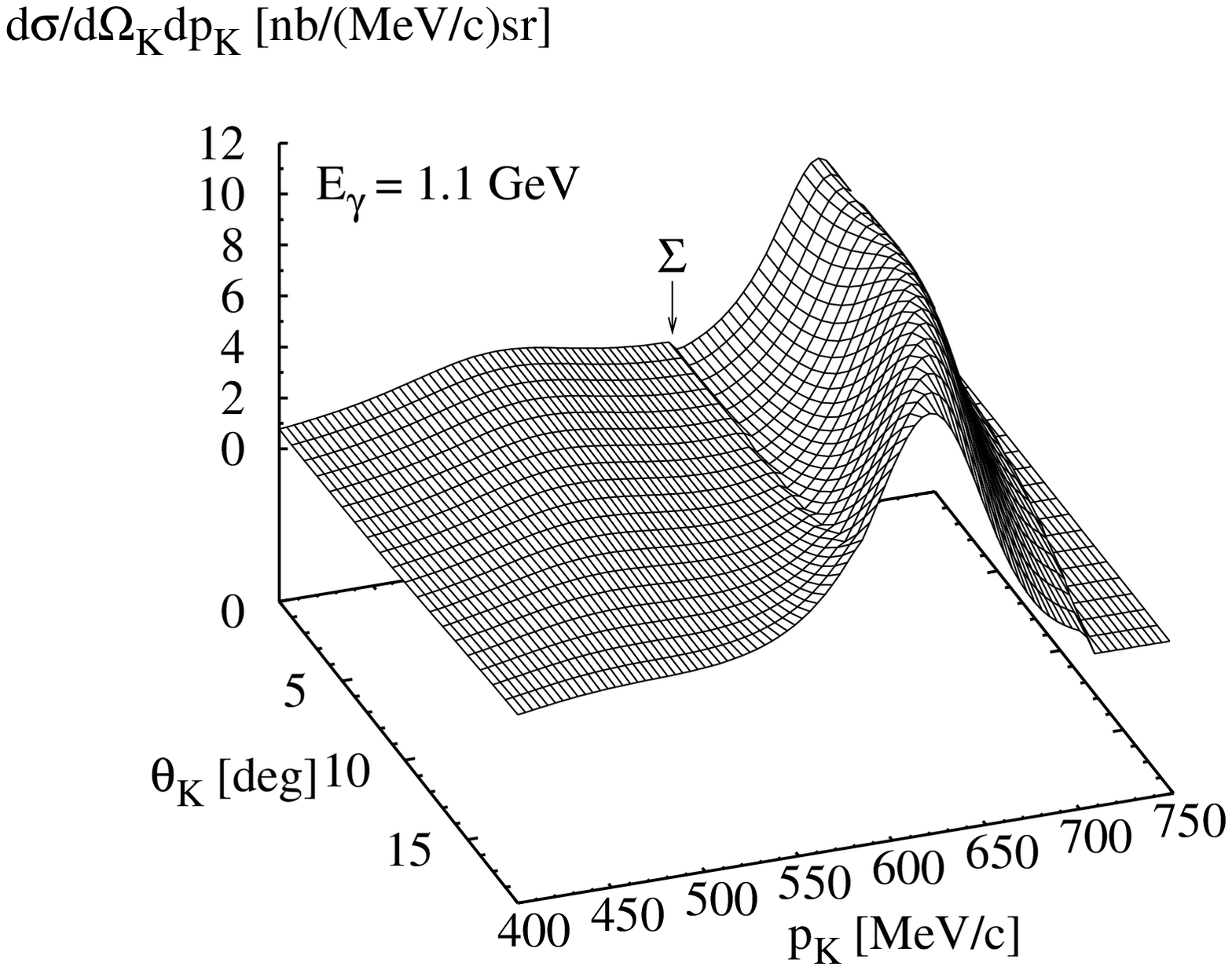}
\end{center}
\caption{Three dimensional plot of $d(\gamma,K^0)YN$
as a function of $\theta_{K}$ and $p_K$, which is obtained
with IA + YN.  The photon energy is $E_{\gamma}$ = 1.1 GeV.
}
\label{fig-three-dim-plot}
\end{figure}
 
Another feature visible in
 Figs.~\ref{fig-qf-exclusive-cross-section-1100}
and \ref{fig-qf-exclusive-cross-section-1300} is
that the cross section of the exclusive process $d(\gamma,K^0\Lambda)p$ 
in QFS kinematics is much larger than in other kinematic regions, 
indicating that measurements in this region will be easier. 
For visualizing the QFS regions over a wide range of $p_{K}$ and
$\theta_{K}$, we provide a three-dimensional plot of
the inclusive process $d(\gamma,K^0)\Lambda p$ obtained with IA + YN
in Fig.~\ref{fig-three-dim-plot}, for the photon energy
$E_{\gamma}$ = 1.1 GeV. 
This figure shows a ridge running along the QFS condition
which  moves to smaller
kaon momentum $p_K$ as the kaon angle $\theta_K$ increases.

\vspace{5mm}
\section{Summary and Conclusion}
\label{conclusions}

We have analyzed $K^0$ photoproduction on the deuteron
$\gamma d \rightarrow KYN$,
investigating the effects of $YN$ and $KN$ rescattering
and the intermediate $\pi N-KY$ rescattering process.
$YN$ rescattering effects are found to be large
in the threshold regions in the inclusive cross section for forward kaon
angle $\theta_{K} = 0^\circ$.
The two models of KAON-MAID for the elementary operator, the original and
the refitted one
to the new SAPHIR data, show a large
variation in the $\Sigma$ QFS region of the inclusive cross sections. Therefore,   
the $d(\gamma,K^0)YN$ experiments can serve as a method to access 
those operators.
In the exclusive processes, the polarization observables
show visible effects of the final-state interactions at
larger hyperon angles.

We have also calculated the exclusive cross section 
for outgoing nucleon momenta $\vec p_{N}$ with
various magnitudes in a fixed direction.
The $YN$ rescattering effects are found to become larger
at forward kaon angles
as the magnitude of $\vec p_{N}$ increases, 
but those are almost negligible at  zero nucleon momentum,
i.e. under QFS kinematics.
We have shown that
the elementary amplitude can be extracted algebraically
from the full amplitude using the impulse approximation when 
FSI effects do not contribute.
We have performed the extraction for the $\vec p_{N}$-values  mentioned
above, and the extracted elementary amplitudes have been 
compared with the free-process ones.
At QFS kinematics the extracted and the free-process amplitudes
agree well. This demonstrates that kaon photoproduction on
the deuteron in the QFS region is suitable for investigating
the elementary process in the neutron channels. 
The exclusive cross sections for the
$d(\gamma,K^0\Lambda)p$ process at the QFS kinematics
are found to be especially large 
suggesting that this region is appropriate for measurements.
We confirm that the region where the cross sections are large
develops close to QFS kinematics and forms a ridge  on the
$\theta_{K}-p_K$ plane. 

\section*{Acknowledgment}
AS would like to thank the Simulation Science Center, Okayama University
of Science, Okayama for the financial support and for the very
kind hospitality during his stay.
This work was supported by the "Academic Frontier"
Project for Private Universities: matching fund
subsidy from MEXT (Ministry of Education, Culture, Sports,
 Science and Technology), Japan.


\end{document}